# SASfit: A comprehensive tool for small-angle scattering data analysis

Authors

**Ingo Breßler[a]\*, Joachim Kohlbrecher[b] and Andreas F. Thünemann[a]**

[a] BAM Federal Institute for Materials Research and Testing, Unter den Eichen 87, Berlin, 12205, Germany

[b] Laboratory for Neutron Scattering, PSI Paul Scherrer Institute, Villigen, CH-5232, Switzerland

Correspondence email: ingo.breßler@bam.de

**Synopsis**   Computer program to perform SAXS and SANS data evaluation.

**Abstract**    Small-angle X-ray and neutron scattering experiments are used in many fields of the life sciences and condensed matter research to obtain answers to questions about the shape and size of nano-sized structures, typically in the range of 1 to 100 nm. It provides good statistics for large numbers of structural units for short measurement times. With the ever-increasing quantity and quality of data acquisition, the value of appropriate tools that are able to extract valuable information is steadily increasing. SASfit has been one of the mature programs for small-angle scattering data analysis available for many years. We describe the basic data processing and analysis work-flow along with recent developments in the SASfit program package (version 0.94.6). They include (i) advanced algorithms for reduction of oversampled data sets (ii) improved confidence assessment in the optimized model parameters and (iii) a flexible plug-in system for custom user-provided models. A scattering function of a mass fractal model of branched polymers in solution is provided as an example for implementing a plug-in. The new SASfit release is available for major platforms such as Windows, Linux and Mac OS X. To facilitate documentation, it includes improved indexed user documentation as well as a web-based wiki for peer collaboration and online videos for introduction of basic usage. The usage of SASfit is illustrated by interpretation of the small-angle X-ray scattering curves of monomodal gold nanoparticles (NIST reference material 8011) and bimodal silica nanoparticles (EU reference material ERM-FD-102).

**Keywords:  SAS, SAXS, SANS, curve fitting, nanotechnology, nanoparticles, polymers.**

## 1. Introduction

With an increasing number of applications for small-angle scattering experiments using high intensity X-rays (SAXS) or neutron beams (SANS), it is becoming increasingly important to have the right tools at hand to help the user to infer valuable information from measurements. Due to its inverse space as well as spanning of several magnitudes of intensity analyzing this type of data is a challenging task. There are several well established programs for model-based SAS data analysis: The *IRENA* package, which comes as an extension to the *IGOR PRO* computing environment, is designed for general analysis and also includes many tools for preliminary data correction (Ilavsky & Jemian, 2009). Scatter is a software for the analysis of nano-and mesoscale SAS (Forster *et al.*, 2010). Furthermore, there is the *ATSAS* project, which consists of a comprehensive set of sophisticated tools primarily intended for SAS data from biological macromolecules (Petoukhov *et al.*, 2012). Recently, the *McSAS* program was also published; it uses a Monte-Carlo algorithm to determine form-free size distributions including uncertainties with known particle shapes (Pauw *et al.*, 2013).

SASfit is a general curve fitting program for analysis of data from small-angle scattering experiments and is mostly used with SAXS or SANS data in the fields of analytical or biological chemistry. In addition, SASfit has been used for the traceable size determination of gold nanoparticles (Meli *et al.*, 2012), polymeric nanoparticles (Gleber *et al.*, 2010) and vesicles (Varga *et al.*, 2014). The program is capable of fitting complex models to several data sets simultaneously. More than 200 particle models are available to set up complex models consisting of several contributions. A flexible plug-in system allows for entirely user defined models.

The central aim of the program is to help users to derive useful information from SAS scattering data by offering basic and advanced curve fitting tasks via an easy-to-use, comprehensive graphical user interface. The software package contains most of the tools required to treat a large range of scientific problems and a large volume of data which is usually acquired on a SAS instrument. Because it is available for the most widely used platforms, i.e. Windows, Linux and MacOS, SASfit is downloaded more than two thousand times per year with an increasing tendency (Sourceforge.net, 2014). According to the download statistics, it has a solid user base in Europe, Brazil, U.S., India and China. The program is distributed under the conditions of the Open Source License GPL which ensures access to its source code and limits redistribution of modifications to the same open license.

The present paper describes important recent improvements of SASfit – which are not self-explanatory – to help the user in solving scientific problems. We provide insight into data import and data reduction options, model configuration and work-flow for curve fitting. Hints on interpretation of the confidence in the fitted parameter values and on data export functions are given. Finally, we explain the programme's capability of producing user-defined SAS model functions via plug-ins.

## 2. Data Import

Typically, SASfit loads three-column ASCII data files for model fitting containing columns for the $q$-vector, intensity and uncertainty of the intensity. It combines them into logical data sets for analysis in a user defined manner. Within a file, anything which can be interpreted as floating point numbers is treated as data. For this reason embedded metadata, produced, for example, by data pre-processing programs like SAXSquant, are ignored by SASfit. This feature enhances convenience of data processing and can be verified easily by opening the data file with an ordinary text editor. If no uncertainties of the intensities are given, SASfit offers an algorithm to estimate them, which is based on the smoothness of the curve.

The menu "Single Data Set" is intended to be used to set up a single dataset for basic analysis. Initially, it loads a data file, creates a scattering curve and launches the "Merge Files" menu, where more data files can be added for different instrumental configurations in order to cover a larger $q$-range. In addition, the "Multiple Data Set" menu allows configuring of several data sets for simultaneous fitting. "Multiple Data Sets" is intended to find a model for a sample measured under different constrains, e.g. a contrast variation or a concentration variation of the same particle to determine the inner structure or the structure factor. Confirming the current set-up in the "Merge Files" menu plots the data set. During data import it is possible to specify a number of lines to skip at the beginning of the file to avoid interpreting embedded metadata as data for analysis. Additionally, it allows customizing the meaning of individual columns in the file. For example, negative values of the intensity can be ignored. A unit conversion at this place simply changes the order of magnitude of $q$-vector values.

Preliminary to data fitting, it is wise to investigate the plot of loaded data and its bars of uncertainties (aka 'errors'). At high $q$, the intensity is typically very low and the uncertainties of the intensities are very high. Accordingly, high-$q$ data contribute little to the overall goodness-of-fit. Therefore, when observing high bars of uncertainties, it is safe to ignore a section of points in the high-$q$ region as they may affect numerical stability and increase calculation time, too. The $q$-range for data can be specified in the overview of merged files for each data file individually.

Further reasons to skip data points at the beginning and end of each loaded file could be that one might have not properly masked the beam stop or that there is still some parasitic scattering from the direct beam, which was not well corrected by background subtraction. These points one often want to skip for the data analysis. Also the last points in a data set sometimes show some artefacts. These intensities are often measured in the corners of a rectangular detector and have not been obtained by an azimuthal average over a very narrow azimuthal angle.

## 2.1. Data Reduction

At this point an important tool for preparing data before analysis is reducing the number of data points of oversampled data sets. This tool helps the user to substantially reduce the time needed for data evaluation, when, for example, 100 instead of 2000 data points are fitted with elaborate model functions. In the "Merge Files" overview, this tool is accessible by a button marked with the associated data file name. Three methods are currently implemented and can be applied according to the preference of the user. Method one and two are intended for quick data evaluation, whereas we recommend the third method for detailed analysis since it maintains most of the original data information.

Method one is simple and straight-forward by thinning out data points by counting according to

$$I_{red,i} = I_{raw,k(i)} \quad \text{with} \quad k(i) = round\left(\frac{i}{ratio}\right) \tag{1}$$

The user specifies a ratio of the original data points to keep. For example, specifying 0.1, as shown in the left-hand panel of Figure 1, ignores 90% of the points for fitting. This rough method is suitable if working with high density SAXS data of several thousand data points and if very fast fitting results are desired to get an initial overview, for example during an experiment.

The second method preserves scattering curve characteristics better than the first. It maintains a user defined distance $\delta_{min}$ between data points by utilizing Pythagoras' theorem in linear or logarithmic two-dimensional space and skips those points which are less far away in accordance with

$$I_{red,i} = I_{raw,k(i)} \quad \text{with} \quad \sqrt{\delta_I(i)^2 + \delta_Q(i)^2} > \delta_{min}. \tag{2}$$

In linear contexts, the $\delta_I(i)$ and $\delta_Q(i)$ are calculated by

$$\begin{aligned}\delta_I(i) &= I_{raw,k(i)} - I_{raw,k(i-1)} \\ \delta_Q(i) &= q_{raw,k(i)} - q_{raw,k(i-1)}\end{aligned} \tag{3}$$

and in logarithmic contexts by

$$\begin{aligned}\delta_I(i) &= \log\left|\frac{I_{raw,k(i)}}{I_{raw,k(i-1)}}\right| \\ \delta_Q(i) &= \log\left|\frac{q_{raw,k(i)}}{q_{raw,k(i-1)}}\right|\end{aligned} \tag{4}$$

The third and most recommended method for data reduction averages neighboured data points locally according to user settings for difference in intensity and width in $q$-space (see right-hand panel of Figure 1). Each local interval $(k,l]$ is determined adaptively so that it contains all points $n$ which fulfil the following condition:

$$\forall n \in (k,l): \frac{|I_k - I_n|}{\delta I_k - \delta I_n} < D_{min} \wedge \frac{|q_k - q_n|}{\bar{q}} < \delta q_{max} \qquad (5)$$

The first parameter $D_{min}$ restricts the intensity difference within an interval in relation to the associated uncertainties (aka 'Error-bars'). Additionally, the maximum width of an interval relative to its position in $q$-space is scaled by the second parameter, $\delta q_{max}$. Both parts of the condition have to be fulfilled by neighbouring data points to produce an interval and thus to allow calculation of an average.

The previous methods do not perform an averaging of data points they just filter them to select only a few which still contain enough information. The last one does a real averaging of data points. If one only uses the criteria for $q$ the effect would be the same than measuring with a lower resolution. This is often justified for oversampled data without any sharp features. In case of a few sharp features the resolution should not be relaxed over the whole $q$-range especially not near the sharp features like a drop in intensity for e.g. the form factor of very unique spheres or sharp increase of intensity close to Bragg peaks. To avoid smearing over such features the first criteria that only data are averaged if they do not differ more than their error bar or a multiple, i.e. fractional of their error bar.

It is important to note that the original data, which is loaded from the input file, is always stored unchanged in the background, and it is also stored for traceability in SASfit project files along with the reduced data. Ignoring parts of the data affects the copy of the data used for numerical analysis only. The selected reduced data for analysis can be changed at any time.

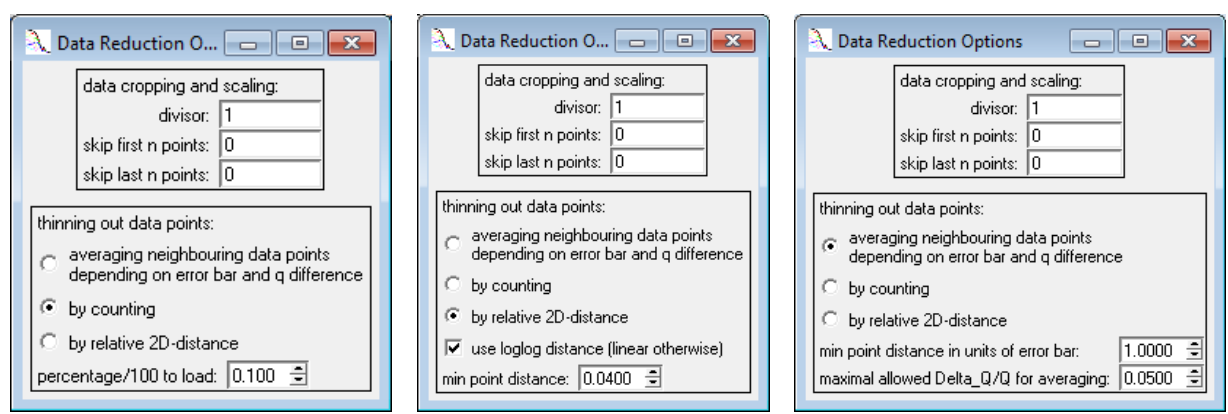

**Figure 1**  Data reduction window providing three different methods for reducing the time required for curve fitting according to the user's preferences. Left-hand figure: Method one skips data points by a count ratio (see eq. (1)). For fast data fitting of $10^3$ data points, a typical percentage/100 to load value of 0.1 is recommended. Middle figure: Method two skips data points within a user defined distance in a two-dimensional space (see eq. (2)). Recommended for fast fitting of data with distinctive curve features. Right-hand figure: Method three averages data points locally and adaptively according to intensity and $q$- spread according to equation (5).

## 3. Model fitting

### 3.1. Model configuration

The main purpose of SASfit is to fit a model of scattering objects to one or more data sets. Selecting "Single Data Set" from the menu bar offers the possibilities of either simulating or fitting data. Both options use the same user interface to set up a model, as shown in Figure 2. The simulate option does not require any input data. Simulate calculates what data of a certain model configuration will look like. A model is set up by choosing one of 200 form factors in the right part of the user interface shown in Figure 2. In this context, the "Sphere" model is the simplest and frequently just as a model, which is first used to describe the scattering curve. It should generally be attempted first if the shape of the particles to be analysed is unknown. Furthermore, the "Sphere" model can be used to evaluate the significance of more complex models by testing experimental data against them.

On the left-hand side of the user interface in Figure 2, one of 20 distribution functions can be selected as a form factor parameter by selecting its *"distr"* column. With a spherical model there is usually a distribution applied to the radius parameter "R". Depending on the scientific field, the Gaussian, Log-Normal and Schulz-Zimm distributions are most often used for polydisperse but monomodal samples. Each distribution consists of at least one parameter controlling the position of its maximum and one parameter controlling its width or FWHM, which defines the degree of polydispersity. The monodisperse distribution implies that all scattering objects have the same size. Also form factors with more than one parameter describing the size of the object are implemented. A simple example would be a spherical shell (core radius and overall size) or a cylinder (diameter and length). SASfit only provides a single distribution of one parameter of the form factor, which however is freely selectable. The user can decide which parameter of the form factor has a distribution.

Right from the beginning of the fit procedure, it is recommended to constrain the optimization algorithm to a physically feasible range of parameter values. If no constraints are applied, there is a risk that either a local fit optimum which does not make sense in the real world, or no solution will be found. For this purpose, next to each model function there is a "Parameter Range" menu, which is only available for single data set and allows the user to set meaningful intervals for each parameter. A quick guide for the user is a short help text, which displays by moving the mouse pointer over the selected function or its parameter. The text shows the implemented formula or a parameter description at the bottom of the window.

Besides form factors and distribution functions SASfit allows consideration of attraction and repulsion of scattering objects by multiplying a structure factor. The latter affects a scattering curve at low $q$-values where the residuum of a fit often shows oscillations if particle interactions are significant. The structure factor can be selected and configured on the second tab "structure factor" of the model configuration view seen in Figure 2. In most cases, the simple "Hard Sphere" structure factor is used. In a basic configuration, its repulsion radius is frequently set slightly larger than the particle radius but maintaining the same order of magnitude, whereas its volume fraction is set to small values of 0.05, for example, at the start of the fitting procedure. Only the monodisperse approximation is a simple multiplication with the structure factor. There are several approximations implemented to include the structure factor. 1) Monodisperse approximation, 2) decoupling approach, 3) local monodisperse approach, 4) partial structure factor, 5) scaling approximation of partial structure factor. For some of the approximations next to the scattering intensity of a particle also the scattering amplitude is needed. Not for all objects the scatter amplitude is known. In this case SASfit might complain. One also has to be careful in case of anisotropic particles, where the structure factor is included via orientational average and size average. Details on the exact formulae are given in the SASfit manual.

To analyse samples containing more than one shape of scattering objects, it is possible to add and manage multiple scattering contributions in a composed model. The contributions of scatterers which do not interact with each other, i.e. have no significant structure factor between different particle species will be summed up for the total intensity, which is then fitted to the measured input data. At the top of the model configuration window in Figure 2, there are buttons to "Add" and "Remove" a contribution as well as switch back and forth to the "Next" and "Previous" scattering contribution. The current contribution is shown in a selection box at the left side of that bar of buttons. Additionally, each contribution can be

1. disabled but not removed by unchecking the "Apply" checkbox, or
2. "Fixed" and thus kept constant during a fit as well as
3. "Subtracted" from the overall model instead of added by default.

When the model and its contributions are configured correctly as desired, the model can be plotted in red colour against the loaded data by using "Apply". The data are plotted as blue dots with error bars.

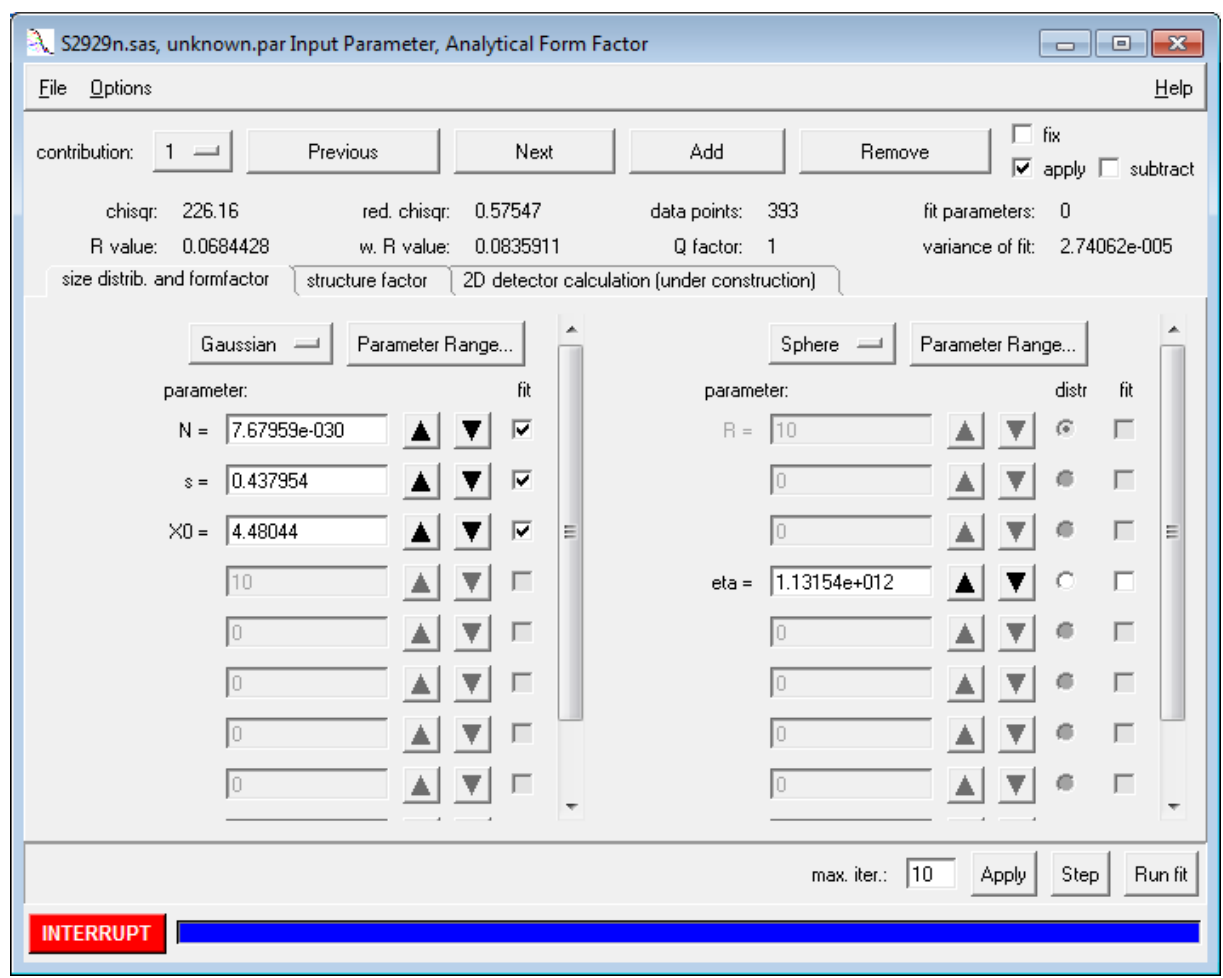

**Figure 2** Basic model configuration consisting of a form factor on the right (in this panel it is a "Sphere" with the scattering contrast of gold nanoparticles "eta" = $1.13 \times 10^{12}$) and a distribution ("distr") of one selected parameter on the left (here it is a "Gaussian" distribution of the radius which has a concentration parameter of "N" = $7.6 \times 10^{-30}$, a width parameter "s" = 0.43 and the mean radius parameter "X0"= 4.48 nm. A Structure factor can be configured on the second tab. Different scattering contributions can be managed by the top row of buttons ("Contributions").

### 3.2. Curve fitting work-flow

Especially with a model consisting of more than one parameter, which is to be optimized against the data, it is advisable to bring the red coloured model curve closer to the data manually by adjusting the initial parameter values. The basic work-flow for fitting small-angle scattering data consists of the following steps as illustrated in Figure 3:

1. The first step of a fit procedure is to match the order of magnitude of fit curve (red) and data intensity (black dots). This can be accomplished by fitting the scaling parameter at the beginning of the curve only. Typically, the first third of the data towards low *q*-values are selected for fitting by dragging with the left mouse button and holding down the *Shift*-key. The selected area is then highlighted by a darker background colour. Usually, the distribution parameter *N* is chosen for this purpose by marking its "fit" checkbox. In a fit, where the data are not measured with absolute intensity values, the "eta" has the same effect as *N* but on an absolute scale "eta" defines the scattering contrast inherent to the sample, whereas the "N" parameter of the distribution denotes the number of scattering objects which were involved in the measurement. Applying "Run fit" finds the optimal value of "N" so that the model and the data curve overlap in the selected area at best and the $\chi^2$ value, defined as

$$\chi(p)^2 = \sum_{i=1}^{N} \left( \frac{I_{exp}(q_i) - I_{mod}(q_i, p)}{\sigma_{exp}(q_i)} \right)^2, \tag{6}$$

is minimized. To plot the model curve with improved scaling over the whole data range, one first clicks on the plot with the left mouse button while holding down the *Shift*–key and then clicks on "Apply".

2. The size of scattering objects is optimized in the second step of curve fitting. Best results are obtained by selecting the central part of the data where, for example, a first local minimum of the curve can be observed. It is important to exclude high intensities at the beginning and increased uncertainties with low intensities at the end of the curve. This time, parameter "*N*" of the distribution function is deselected for the fit, but the radius "X0" (in the Gaussian distribution) is selected instead. This optimizes the location of the maximum of the distribution. Running the fit improves how the model curve reproduces local minima in the data.

3. Fitting both the scaling parameter and the size parameter at the same time over the first two thirds of the data further improves the overall quality of the fit.

4. Some slight mismatch in the central part of the scattering curve can be optimized by fitting the particle radius together with the distribution width parameter "s" (width of the Gaussian distribution in the example). With increasing value of the size distribution width parameter, the model curve becomes smoother since it represents a broader size range of scattering objects. Because of this smoothing effect, it is increasingly difficult to distinguish different shapes with a large polydispersity and thus the significance of a certain shape may decrease because another shape could fit the data equally well. Therefore, care has to be taken in interpretation of broad size distributions which are larger than 20 % of the mean value. A smoothing effect can also be observed using instrumental smearing, which is based on the fact that the incident beam contains a distribution of wavelengths and it is therefore not monochromatic. Wavelength effects are relevant in SANS, but of minor importance in SAXS.

Usually, the previous steps 1 - 3 are repeated until a good overlap of the model curve and the data is obtained. Polydispersity is applied as a last step due to its strong smoothing effect.

A simple and typical example for applying this fit procedure is the determination of the mean radius of spherical particles, the width of their radii distribution and the particle number concentration. We measured a dispersion of gold nanoparticles of the NIST reference material RM-8011 (De Temmerman *et al.*, 2014) (Kaiser & Watters, 2007) for 30 minutes and processed the data to absolute intensity using water as a primary standard as described by Orthaber (Orthaber *et al.*, 2000). The resulting data and a curve fit using a model of spheres with Gaussian size distribution is shown in Figure 4 (black dots and red solid curves, respectively). The uncertainties of the intensity values are displayed as vertical blue lines. The fit parameters in this example are values for the particle concentration $N$, the mean particle radius $X_0$ of the assumed Gaussian size distribution and the width of the size distribution $s$. Note that numerous size distributions are provided including the frequently used Schulz-Zimm (Flory) and Log-normal distribution. It is the user's choice to select the most appropriate one. We recommend utilizing first the Gaussian size distribution if no evidence is available to prefer a type of size distribution a priori, e.g. from other methods like electron microscopy. The best fit values for the mean radius in our example is $X_0 = (4.48 \pm 0.05)$ nm and the width of the size distribution is $s = (0.44 \pm 0.05)$ nm. Note that the uncertainties of the values are only the uncertainty contributions from the fit. These uncertainties have been utilized for determination of the combined standard uncertainties from all input quantities (Meli *et al.*, 2012). But such is a tedious procedure which is beyond the scope of this report. As a rule of thumb the uncertainty of the size parameters from SASfit is typically of the same order of magnitude than the combined standard uncertainties. The factor necessary to convert the $N$–value to a particle concentration in number of particles per cm$^3$ depends on the units used for absolute intensity, scattering vector and scattering length density. Here, this conversion factor is $10^{42}$ since the corresponding units used were cm$^{-1}$, nm$^{-1}$, and cm$^{-2}$. Therefore, in our example, the $N$–value of $(7.68 \pm 0.28)\times 10^{-30}$ corresponds to a particle number concentration of $(7.68 \pm 0.28)\times 10^{12}$ cm$^{-3}$, or in molar concentration it is $(1.28 \pm 0.05)\times 10^{-7}$ mol L$^{-1}$. We recommended to check the plausibility of the particle number concentration and calculate their mass fraction $\phi_m$. For a Gaussian size distribution the mass fraction is $\phi_m = N\varrho\langle V\rangle = N\varrho \times \frac{4}{3}\pi X_0^3 \left(1 + 3\left(\frac{s}{X_0}\right)^3\right)$, where $\varrho$ is the density and $\langle V\rangle$ the mean particle volume. Assuming that the gold particles in our example have the same density of 19.300 g cm$^{-3}$ as the bulk material, we calculated a mass fraction of $(57.37 \pm 2.09)$ µg g$^{-1}$. This value is in reasonable agreement with the value of $(51.56 \pm 0.23)$ µg g$^{-1}$ provided by NIST (Kaiser & Watters, 2007), which was determined by inductively-coupled plasma optical emission spectrometry (ICP-OES). It should be noted that the uncertainty of intensity measurements is typically in the order of 5 % (Orthaber *et al.*, 2000) and therefore the real uncertainty of $N$ must be at least of the same magnitude, i.e. larger as derived from the curve fit alone.

In general, it is an open question what a *"good"* fit means in the context of small-angle scattering: Concerning $\chi^2$ which is subject to the underlying optimization, it depends heavily on the quality of the data and how well the associated uncertainties were estimated. If the programme considers the uncertainties to be too large, a $\chi^2$ of less than 1 is found, whereas underrated uncertainties often give a $\chi^2$ greater than 1. But in the theory of experimental data acquisition, the resulting optimal reduced $\chi^2$ is supposed to be exactly 1. There are several statistical measures which aim to provide an alternative approach to the goodness of fit in order to compare curve fit results.

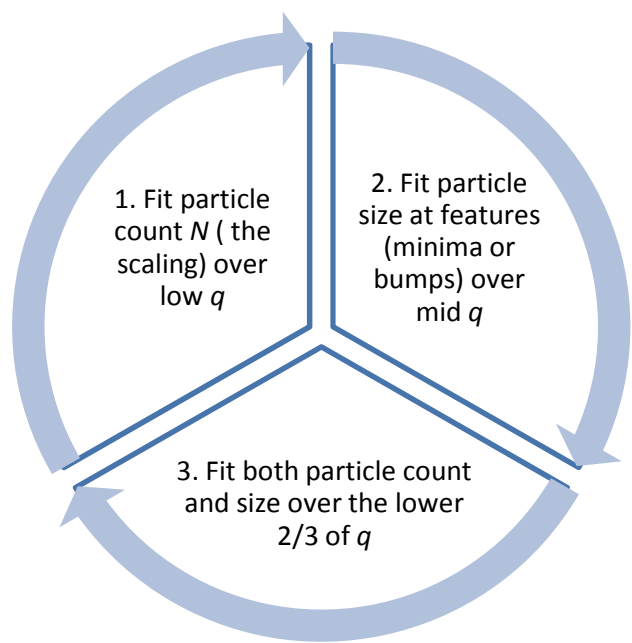

**Figure 3** The basic curve fitting work-flow of a three step circle is typically recommended for using SASfit.

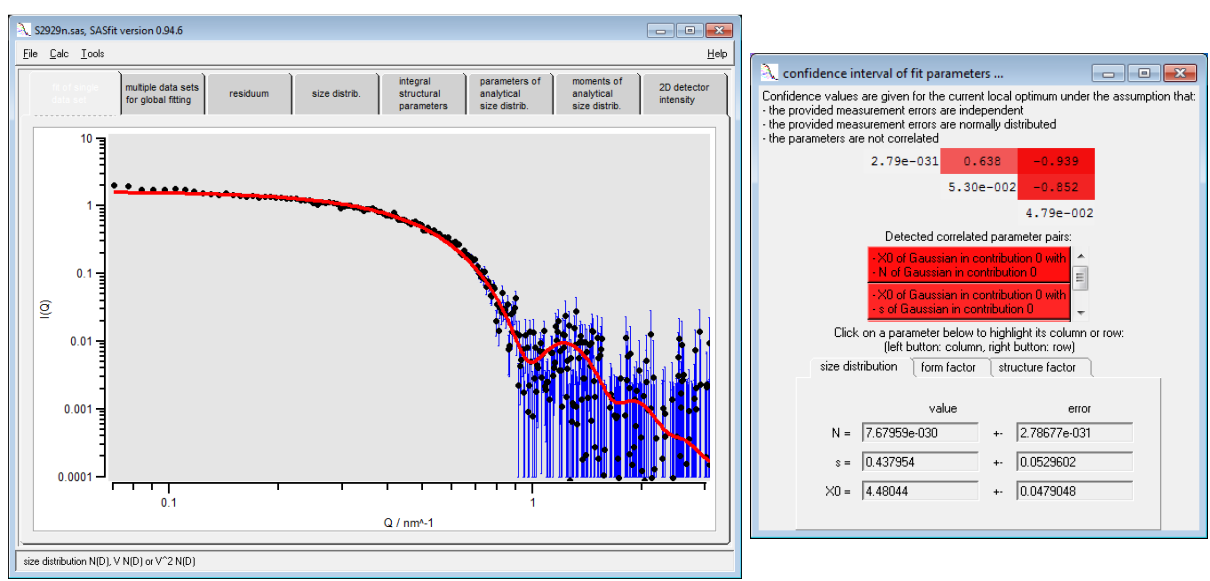

**Figure 4** Curve fit of a single data set. Left-hand figure: SAXS data of gold nanoparticles of the NIST reference material RM-8011 (De Temmerman *et al.*, 2014) (Kaiser & Watters, 2007) and a curve fit using a model of spheres with Gaussian size distribution (black dots and red solid curves, respectively). The uncertainties of the intensity values are displayed as vertical blue lines. The fit parameters are values for the particle concentration *N*, mean particle radius $X_0$ and width of the size distribution *s*. The conversion factor of the *N*–value to a concentration value in number of particles per cm$^3$ depends on the units used for absolute intensity. This conversion factor is $10^{42}$ if the corresponding units are cm$^{-1}$, nm$^{-1}$, and cm$^{-2}$. Therefore, the concentration of particles is calculated as $c = (7.68 \pm 0.28) \times 10^{12}$ cm$^{-3}$, which is $(1.28 \pm 0.05) \times 10^{-7}$ mol L$^{-1}$. The estimate for the mean radius is $X_0 = (4.48 \pm 0.05)$ nm and the width of the size distribution is $s = (0.44 \pm 0.05)$ nm. Right-hand figure: Covariance matrix and confidence intervals of fit parameters to assist in revealing correlated parameters. The concentration of the gold nanoparticles given in the certification report is $(51.56 \pm 0.23)$ µg/g which is only $2.67 \times 10^{-4}$ volume %.

### 3.3. Fit quality

Information on the results of the fits is provided for the user. After each fit SASfit displays measures in the model configuration window, which serve as an indicator of the fit quality (see Figure 2). The two basic measures are the "chisqr" value $\chi^2$ and the "reduced chisqr" value $\chi^2_{red} = \chi^2/(N - M)$, where *N* is the number of data points used for the fit and *M* is the number of parameters of the fit model. The $\chi^2_{red}$ provides a measure of fit quality across data sets and model configurations. The main question for evaluating the quality of a fit is about the relevance of the data with respect to the model and its parameters: Would it provide the same fit quality with the same values for another data set, possibly random data? The $Q_{\text{factor}}$, defined as

$$Q_{factor} = Q\left(\frac{N-M}{2}, \frac{\chi^2}{2}\right) = \frac{\Gamma\left(\frac{N-M}{2}, \frac{\chi^2}{2}\right)}{\Gamma\left(\frac{N-M}{2}\right)} \text{ with } \Gamma(a,x) = \int_x^\infty t^{a-1} e^{-t}\, dt, \tag{7}$$

provides valuable information for solving this question, namely the probability that a random set of N data points would produce an equal or higher value of $\chi^2$ with the same model configuration. For a fit of good quality, its value is supposed to be in the range of 0.01 and 0.5 along with a $\chi^2_{red}$ value close to one.

In analogy to the *R*-factor in crystallography (Hamilton, 1965, IUCr, 2008), SASfit provides an "R-value" as quality criteria of a model in data analysis results:

$$R = \frac{\sum_{i=1}^{N} \left| |I_{exp}(q_i)| - |I_{mod}(q_i)| \right|}{\sum_{i=1}^{N} |I_{exp}(q_i)|} \tag{8}$$

If the model perfectly fits the measured data, the $R$ is zero, whereas it approaches infinity if the model configuration diverges from the data. A value of 0.1 is frequently used as the threshold for an acceptable fit quality. It is especially important to realize that $R$ is only a measure of precision, and that it is not able to measure accuracy. Cases of data situations and model combinations that would be reported as false positives or negatives by the value of $R$ are conceivable. Since the function being minimized is weighted by the uncertainties of the measured data (as can be seen in eq. (9)), there is a weighted R-value "w. R value" provided, which takes those estimates of precision into account by

$$R_w = \sqrt{\frac{\sum_{i=1}^{N}\left(\frac{|I_{exp}(q_i)| - |I_{mod}(q_i)|}{\sigma_{exp}(q_i)}\right)^2}{\sum_{i=1}^{N}\frac{I_{exp}^2(q_i)}{\sigma_{exp}^2(q_i)}}}. \qquad (9)$$

### 3.4. Confidence in fitted parameter values

In addition to the above mentioned values that characterize of the overall fit quality, SASfit provides the user with confidence intervals for the fitted parameters and outputs the internal covariance matrix to support the identification of highly dependent parameters. The respective "confidence intervals of fit parameteres" shown in Figure 4 is accessible via the options menu of the model configuration window. In order to find optimal model parameters SASfit uses the Levenberg-Marquardt algorithm (Levenberg, 1944) to minimize the $\chi^2$ function (see eq. (6)). During optimization the algorithm approximates its Hessian matrix, which consists of all partial second derivatives according to the parameters being fitted. The inverse of the Hessian matrix is the approximated formal covariance matrix $\mathbf{C}$ for the fit. The square-root of diagonal elements $\sqrt{C_{jj}}$ gives the standard deviation $\sigma = \delta p_j$ of the best-fit parameter $p_j$, which holds only under the assumption that measurement errors are independent and normally distributed as well as that the parameters are not correlated to each other. Since there are no means to verify those assumptions, the given confidence intervals have to be interpreted with care. Note that SASfit provides the standard deviation of the fit parameters from which confidence intervals can be derived according, for example, to the Guide to the Expression of Uncertainty in Measurements (GUM) (Metrology, 2008). In order to assist the user in assessing the last assumption, the correlation coefficient $r_{jk}$ of every pair of fit parameters is shown in the upper triangular matrix in shades of red depending on their degree of correlation. For two parameters $p_j$ and $p_k$ being optimized the correlation coefficient $r_{jk}$ is given by

$$r_{jk} = \frac{C_{jk}}{\sqrt{C_{jj}C_{kk}}} \qquad (10)$$

For uncorrelated parameters, $r_{jk}$ is expected to have a value close to zero, whereas for strongly correlated parameters $|r_{jk}|$ approaches one. When two parameters are strongly correlated it can happen that they both are unphysically large or small. In such cases one has either to rewrite the form factor with less parameters so that the two correlated parameters are combined to only one, which is often not possible, or one simply fixes one parameter to a value which one knows in the best case from another technique. Another strategy would be to think about a SAS experiment, like contrast variation, etc. to decouple the two strongly correlated parameters. There is one row and one column associated with each parameter being optimized. They can be highlighted by clicking on a parameter entry in the lower half of the window. By selecting the row and column of two different fit parameters, their correlation coefficient $r_{jk}$ at the position of their common matrix element is highlighted.

### 3.5. Data export

Parameters and confidence values of the latest fit can be found under the "parameters of analytical size distrib" tab in the main window. It shows all of the configured model functions along with their parameters as text for easy export, and in the context menu (right-click) it offers to write them to a semicolon separated text file. The semicolon separated text output consists of three columns for the size distribution followed by three columns at the centre making up the form factor settings as well as three columns at the end for the structure factor. In addition to the model parameters, the moments and other statistics calculated for the distribution function are also given. Those values can be found under the tab "moments of analytical size distrib" and can be exported in the manner described above.

### 4. Batch processing

Once a model has been configured, it can be used for processing a batch of data files under "Options" → "run batch", as shown in Figure 5. Holding the mouse pointer over the pattern input field reveals a short pop-up help text field on the pattern syntax for file selection. SASfit allows filtering of data file names from a user-defined input directory for model dependent analysis as well as for model independent analysis.

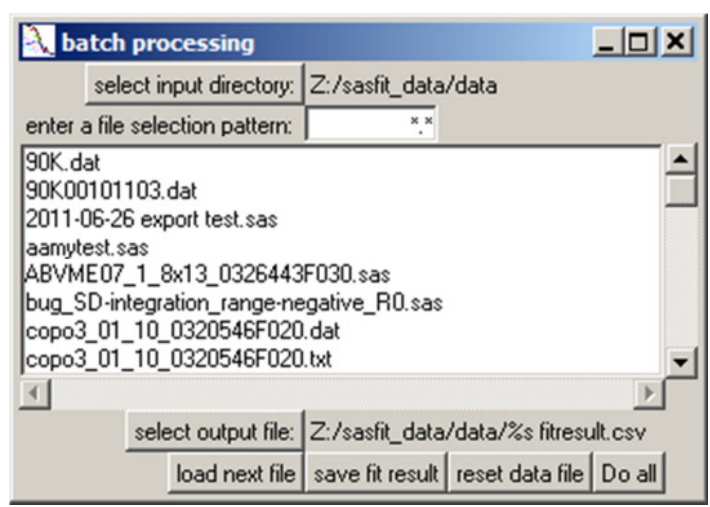

**Figure 5**  Panel for selection of data files for batch analysis and individual output file.

## 5. Custom model functions as plug-ins

In addition to the large library of existing model functions for form factors, structure factors and size distributions, SASfit features a flexible plug-in system, which allows for custom model functions and provides everything to enable users to write their own custom form factor and structure factor functions in the C programming language.

**Plug-in concept**

In SASfit a plug-in is a container for model functions. It may contain an arbitrary number of form factors and structure factors. Both types are supported within a single plug-in at the same time, but it is recommended to use a plug-in for grouping model functions of a similar kind. In this way, for all model functions of a plug-in, a common set of internal helper routines not accessible publicly can be created and used. SASfit plug-ins can be exchanged freely between different SASfit installations even in binary form if the PC platforms and architectures are compatible to each other. To create new customized plug-ins it is strongly recommended to build SASfit from its source code first. In this way, the build environment is verified to work correctly and also the plug-in system compatibility is assured.

**Retrieving the source code**

The latest source code of SASfit including a history of all changes can be browsed and downloaded on the code hosting page of the project (http://sourceforge.net/projects/sasfit/). There are two options to get the most recent source code files: (a) By using the distributed version control system (DVCS) Mercurial to 'clone' the project repository locally. This requires a third party client program for Mercurial being installed but it simplifies the effort of updating to a new version into a single mouse-click. Alternatively, (b) the history view of past 'commits' on the code hosting page provides a download link for the complete source code of the selected version called a "Snapshot". Be sure to select the version tagged by "tip" which always points to the latest version automatically. The technical details on the required build environment and the specific instructions for building the SASfit program on a specific platform can be found in the online documentation (http://sourceforge.net/projects/sasfit/).

**Creating a new plug-in**

After successfully building SASfit, the program is run directly from its source code directory in order to create a new empty plug-in template containing a directory structure of source code skeleton files. For this purpose, SASfit provides a plug-in guide shown in Figure 6. It can be found under the main menu "Tools" → "create new plug-in" and lets the user define the set-up of a new plug-in function. It requires a new plug-in name to be provided by the user to differentiate it from existing plug-ins while at least one function has to be configured including a name under which it can be found in the model selection menu finally. Additionally, the plug-in guide expects the required parameters of each function to be defined. It is important to clarify the numerical implementation of the desired model function beforehand and thus knowing the specific parameters needed. Because modifying existing model functions cannot be done easily by the user, it is recommended to recreate a plug-in with the respective functions in case modifications are required.

Initially, newly created plug-in templates already contain the configured model functions but they lack any functionality and evaluate to constant zero. This ensures that the plug-in can be build right from the beginning by issuing the previously used build commands again. It will build only those source code files which are new or changed since the last run. In this case it is supposed to build the newly created empty plug-in only and add its binary files to the appropriate location automatically. To verify that the plug-in was built correctly SASfit has to be started again and the new plug-in will be listed in the appropriate model selection list under "by plug-ins".

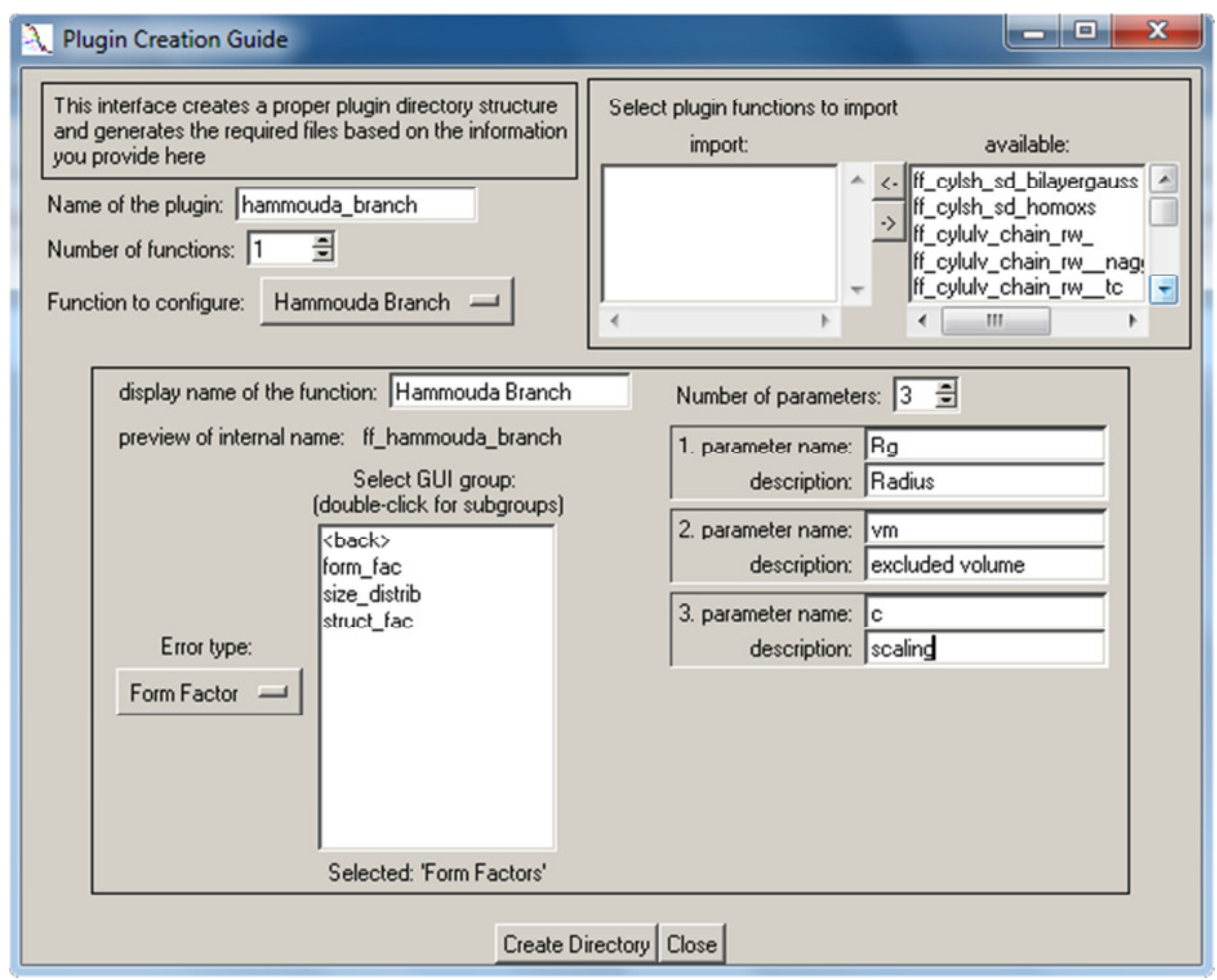

**Figure 6** User interface for creating a new plug-in template consisting of user-defined model functions, filled out according to the branched polymer example plug-in.

**Branched polymer plug-in function example**

Once the initial build of the new plug-in succeeded it can be populated with the desired model implementation. An example is presented in the following by implementing a single-polymer form factor for branched polymers formulated by Boualem Hammouda (Hammouda, 2012).

By using a (mass) fractal model for the minimum path corresponding to the main chain backbone of the polymer the form factor is described by

$$P_B(Q) = \frac{1}{Norm} 2 \int_0^1 dx (1-x) x^{c-1} \exp[-U_B x^{2\nu}]$$

with the normalization factor being defined by

$$Norm = 2 \int_0^1 dx (1-x) x^{c-1} = \frac{2}{c(c+1)}$$

and the scattering variable $U_B$ is expressed in terms of the radii of gyration $R_g$:

$$U_B = Q^2 R_g^2 \frac{(2\nu + c)(2\nu + c + 1)}{6}$$

with a change of variable in $t = U_B x^{2\nu}$ and $dt = 2\nu\, U_B x^{2\nu-1} dx$ the integral $P_B(Q)$ evaluates to

$$P_B(Q) = \frac{1}{Norm} \left[ \frac{1}{\nu\, U_B^{c/2\nu}} \gamma\left(\frac{c}{2\nu}, U_B\right) - \frac{1}{\nu\, U_B^{(c+1)/2\nu}} \gamma\left(\frac{(c+1)}{2\nu}, U_B\right) \right]$$

The remaining variables $\nu$ for the excluded volume and $c$ for the scaling factor become parameters of the model function next to the radii of gyration $R_g$. This formulation of the form factor translates into the following source code of the respective model function in a SASfit plug-in:

```
01 scalar sasfit_ff_hammouda_branch(scalar q, sasfit_param * param)
02 {
03     scalar ub, norm_inv;
04     scalar (*gamma) (scalar, scalar);
05
06     SASFIT_ASSERT_PTR(param); // assert pointer param is valid
07
08     // modify conditions to your needs
09     SASFIT_CHECK_COND1((q < 0.0), param, "q(%lg) < 0", q);
10     SASFIT_CHECK_COND1((RG < 0.0), param, "Rg(%lg) < 0", RG);
11     SASFIT_CHECK_COND1((VM < 0.0), param, "vm(%lg) < 0", VM);
12     SASFIT_CHECK_COND1((C < 0.0), param, "c(%lg) < 0", C);
13
14     // insert your code here
15     ub = q*q * RG*RG * (2.*VM + C) * (2.*VM + C + 1.) / 6.;
16     norm_inv = .5 * (C*C + C);
17     gamma = gsl_sf_gamma_inc_P;
18     return ( gamma(.5* C     /VM, ub) / (VM*pow(ub, .5* C     /VM))
19            - gamma(.5*(C+1.)/VM, ub) / (VM*pow(ub, .5*(C+1.)/VM))
20           ) * norm_inv;
21 }
```

The function signature in line 1 was created by the plug-in guide along with the mandatory verification of input parameters in line 6. This function is evaluated for every individual $Q$ value of the scattering vector provided in the first argument "scalar q". Access to predefined input parameters of the model function is provided by automatically generated variables in upper case: "RG", "VM" and "C" along with range checks on them in line 9-12 which were adjusted to a reasonable domain. Each range check consists of different parts: The first part is the condition which will raise an error, for example in line 9 the scattering vector Q being smaller than zero. The next part is the name of the common parameter structure which is in most cases "param". All following parts of a check define an error message to be forwarded to the user. At the beginning of each model function all variables which will be used are declared. Line 3 in this example declares two floating point variables which will be defined later while line 4 declares a short-cut name of a function which expects two input values. In line 17 it is set to a specific gamma function provided by the GNU Scientific Library (GSL). The model function defined by the formula of Hammouda for branched polymers is actually implemented on lines 15 to 20 with the scattering variable $U_B$ defined on line 15, the inversion of the normalization factor on line 16 replaces two divisions by multiplications in the final formula on line 20. This code replaces the automatically generated template source code of the function "sasfit_ff_hammouda_branch()" in file "sasfit_ff_hammouda_branch.c" of the plug-in template which was generated by filling out the SASfit plug-in guide as shown in Figure 6.

As demonstrated in the example, model functions in SASfit can make use of any function in the GNU Scientific Library (Galassi & Gough, 2009) but may also use a large set of predefined mathematical functions provided by SASfit directly. For example, a convenient wrapper "sasfit_integrate()" which simplifies usage of GSL integration routines by managing workspace memory in the background. Additionally, custom routines can make use of model functions defined in other plug-ins by declaring to import them during configuration with the plug-in guide. More information on plug-ins in SASfit an extensive guide on how to start writing custom models for SASfit on the Windows, Linux or MacOS platform as well as video guides can be found online (http://sasfit.sf.net/manual/Overview:_Plugins).

## 6. Example: Characterization of a bimodal silica particle size distribution

The interpretation of multimodal size distributions of nanoparticles is a demanding typical SASfit application. The procedure is explained in the following by interpreting the scattering pattern of a bimodal size distribution of silica nanoparticles in aqueous solution. Recently, a suitable particle mixture was released as a certified reference material denoted ERM-FD-102 (Kestens & Roebben, 2014) which is commercially available as a European Reference Material. The intended use of ERM-FD-102 is the quality control and assessment of performance of nanoparticle size analysis methods, including SAXS. We have chosen ERM-FD-102 in order to allow all SASfit users to check our results easily and to verify the appropriate use of SASfit. A sample volume of 20 μL was measured as received for 30 min on a commercial SAXS instrument and its scattering intensity was converted to absolute scale using water as primary standard according to the procedure described by Orthaber et al. (Orthaber et al., 2000) and it was verified using a measurement of bovine serum albumin (Mylonas & Svergun, 2007). The resultant scattering curve with data in the range of $q_{min}$ = 0.057 nm$^{-1}$ to $q_{max}$ = 3.0 nm$^{-1}$ is shown in Figure 7. In a first step of data evaluation we calculated the scattering contrast "eta" between the silica particles and the solvent. For this purpose we used the scattering length density calculator, which is available at the "Tools" menu entry. The silica particle scattering length density was calculated as 1.962 10$^{11}$ cm$^{-2}$ by inserting the stoichiometry of silica SiO$_2$, the density of (2.29 ± 0.01) g cm$^{-3}$ (Finsy et al., 1985) and the copper tube X-ray energy of 8.042 keV (see Figure 7). Thus after subtracting the scattering length density of water (9.45 × 10$^{10}$ cm$^2$) , which was calculated in the same way, we obtained a silica particle scattering contrast of 1.017 × 10$^{11}$ cm$^{-2}$. Next, a sphere model for the particles' form factor with a Gaussian size distribution was chosen at the "Calc" – "Single Data set" – "fit" menu entry. Therein the value of "eta" was inserted for "contribution 1" and "contribution 2" as a fixed parameter. Then we performed the fitting procedure described in the curve fitting workflow section (see also Figure 3). The resultant best fit curve is shown together with the data points in Figure 7 (red solid curve and points, respectively). The corresponding best fit values are displayed in the fit panels for the particle size contribution 1 and 2, respectively (lower row of Figure 7). The best fit values of the parameters and estimates of their uncertainties are displayed when clicking the button "parameters of analytical size distribution". All values can be copied to the clip-board via the right mouse button or saved in a file for further use. The parameters of the ERM-FD-102 sample were transferred to Table 1. The estimate for the mean radius of silica particle class A is $X_0$ = (8.52 ± 0.04) nm and (37.65 ± 3.30) nm for class B. These values are in good agreement with the number-weighted modal area-equivalent radii of (9.1 ± 0.8) nm and (42.0 ± 1.1) nm obtained by transmission and scanning electron microscopy (Kestens & Roebben, 2014). On a first sight it is surprising that the uncertainty of the mean radius is much larger for the larger particles than for the smaller ones. But when taking into account that the largest dimension, which can be "seen" is about $\pi/q_{min}$ = 56 nm it becomes clear that the particles of class B are at the upper size resolution limit of the measurement. In contrast, the radii of class B particles are far away from the upper resolution limit and also from the low resolution limit of $\pi/q_{max}$ = 1 nm. Accordingly the uncertainty of the radii of class B becomes relatively large in comparison to the particles of class A. The width of the size distributions of class A and B are s = (2.00 ± 0.03) nm and (8.29 ± 3.04) nm, respectively, which are typical values for commercial silica particles. Also for s, the uncertainty for class B is larger than for class A for the same reason as for $X_0$. It should be noted that SASfit allows an estimation of the number-weighted size distribution of the particles, which could be relevant for the data interpretation and must always be kept in mind. Number-weighted size distributions are important for the characterization of nanomaterials which are defined by the European Commission as *"A natural, incidental or manufactured material containing particles, in an unbound state or as an aggregate or as an agglomerate and where, for 50 % or more of the particles in the number size distribution, one or more external dimensions is in the size range 1 nm - 100 nm"* (Potočnik, 2011).

Here, SASfit provides direct access to an estimate of number-weighted size distributions of nanoparticles. The implemented formula for curve fitting of spheres, $I_{sphere}(q, R, \Delta\eta)$, with Gaussian number-weighted size distribution, $\text{Gauss}(R, N, \sigma, R_0)$, is

$$I_{SASfit}(q) = \int_0^\infty \text{Gauss}(R, N, \sigma, R_0) \, I_{sphere}(q, R, \Delta\eta) \, dR$$

where the Gaussian size distribution is defined as

$$\text{Gauss}(R, N, \sigma, R_0) = N \left[ \sqrt{\pi/2} \, \sigma \left(1 + \text{erf}\left(R_0/(\sqrt{2}\,\sigma)\right)\right) \right]^{-1} e^{\frac{-(R-R_0)^2}{2\sigma^2}}$$

and the scattering of a sphere is given by

$$I(q, R, \Delta\eta) = \left[ \frac{4}{3}\pi R^3 \, \Delta\eta \left( 3 \frac{\sin(qR) - qR\cos(qR)}{(qR)^3} \right) \right]^2$$

The approach to estimate the number-weighted distribution is only useful if the type of distribution is relatively narrow, typically smaller than 20 % relative width, or it can be reasonably estimated before fitting in cases of broad size distributions. In contrast, the recently published Monte-Carlo approach for analysis of SAS data provides good estimates of volume-weighted size distribution but is much less suited for number-weighted ones (Pauw *et al.*, 2013). In SASfit the size distribution is the number density as long as the form factor is expressed in terms of a size. The form factor contains always the volume information. If needed, one could easily implement a form factor of spheres with input parameters volume and contrast and then have a Gaussian distribution of volumes. As described for the gold nanoparticles above the fitted *N*–values of class A and B particles were converted to particle number concentrations of $(1.02 \pm 0.01) \times 10^{15}$ cm$^{-3}$ and $(6.51 \pm 1.48) \times 10^{-11}$ cm$^{-3}$, respectively. Molar concentrations were $(1.69 \pm 0.02) \times 10^{-6}$ mol L$^{-1}$ and $(1.08 \pm 0.25) \times 10^{-9}$ mol L$^{-1}$. Therefore, the number ratio of small to large particles $N_1/N_2$ is 1567 ± 371. We also calculated the mass fractions assuming that the silica particles in our example have a density of 2.29 g cm$^{-3}$ (Finsy *et al.*, 1985) resulting in $\phi_{m,1}$ = (7.05 ± 0.07) mg g$^{-1}$ for class A and $\phi_{m,2}$ = (0.38 ± 0.09) mg g$^{-1}$ for class B. Based on the composition data given in the certification report of ERM-FD-102 (Kestens & Roebben, 2014) we calculated $\phi_{m,1}$ = 8.33 mg g$^{-1}$ and $\phi_{m,2}$ = 0.42 mg g$^{-1}$. From these values the mass ratio derived from SASfit is $\phi_{m,1}/\phi_{m,2}$ = 18.5 ± 4.6 and from the certification report it follows that $\phi_{m,1}/\phi_{m,2}$ = 19.8. We conclude that precision and accuracy of the SASfit parameters and thereof derived values are in good agreement with the reported values of the silica reference material.

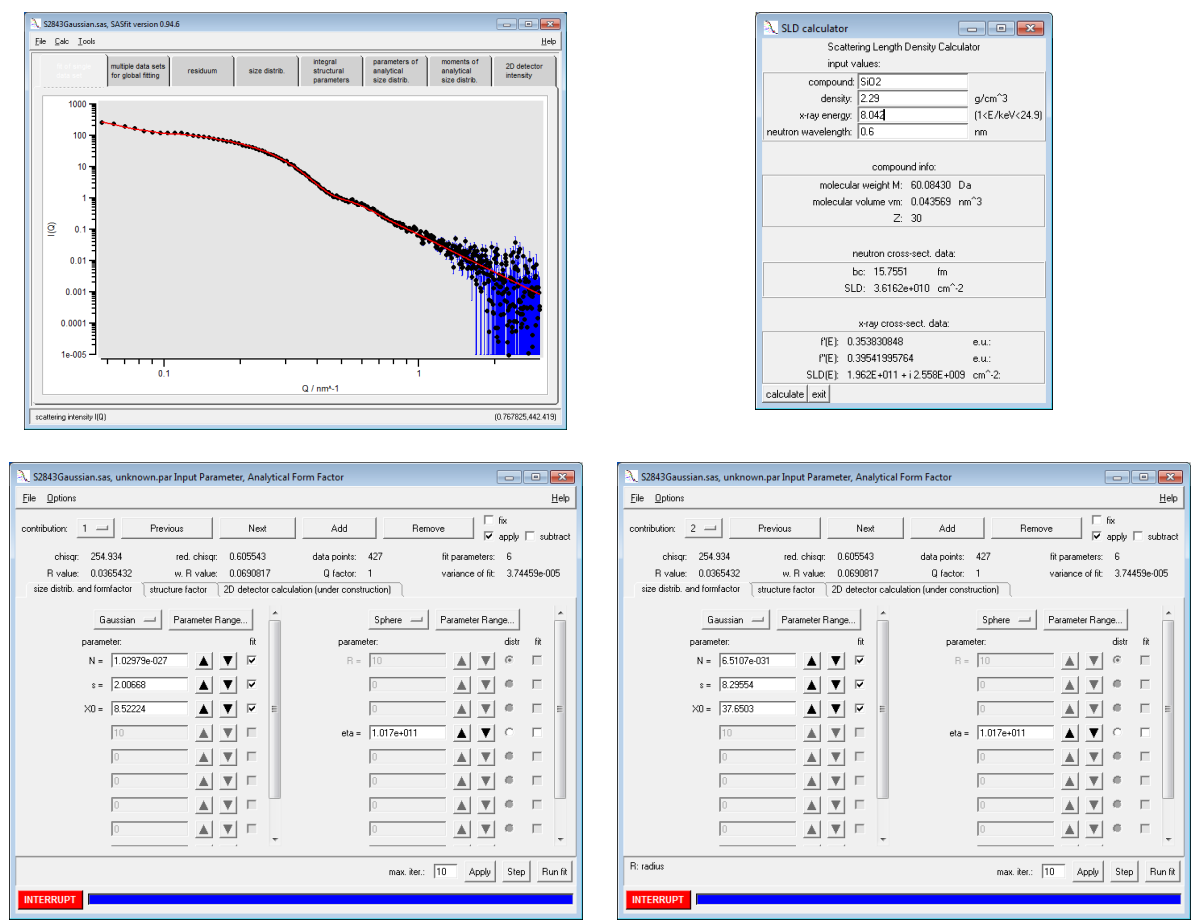

**Figure 7** Upper left-hand figure: SAXS data of bimodal silica nanoparticles (European reference material ERM-FD-102) and a curve fit using a model of spheres with Gaussian size distribution (black dots and red solid curves, respectively). The uncertainties of the intensity values are displayed as vertical blue lines. Upper right-hand figure: The scattering length density calculator provides a scattering length density of $1.962 \times 10^{11}$ cm$^{-2}$ for $SiO_2$ particles with a density of 2.29 g cm$^{-3}$. Lower figures: Panels of the sphere form factor with Gaussian size distribution for the small particles (contribution 1, left-hand) and large particles (contribution 2, right-hand).

**Table 1** Parameters of silica nanoparticles ERM-FD-102 fitted with a bimodal Gaussian size distribution. The population of small particles are labeled as "Particle class A" and that of the large particles as "Particle class B" in accordance with the ERM-FD-102 certification report (Kestens & Roebben, 2014). Fit parameters are the particle number $N$, the mean radius $X_0$, and the width of the size distribution $s$. Values of the mass fraction of the particles $\phi_m$ are given as derived from the SASfit parameter $N$ and calculated from the data given in the certification report.

| Parameter | Particle class A | Particle class B |
| --- | --- | --- |

|  | Contribution 1 | Contribution 2 |
|---|---|---|
| $N$ | $(1.02 \pm 0.01)\, 10^{15}$ cm$^{-3}$ | $(6.51 \pm 1.48)\, 10^{11}$ cm$^{-3}$ |
|  | $(1.69 \pm 0.02)\, 10^{-6}$ mol L$^{-1}$ | $(1.08 \pm 0.25)\, 10^{-9}$ mol L$^{-1}$ |
| $\phi_m$ (SASfit) | $(7.05 \pm 0.07)$ mg g$^{-1}$ | $(0.38 \pm 0.09)$ mg g$^{-1}$ |
| $\phi_m$ (certification report)[a] | 8.33 mg g$^{-1}$ | 0.42 mg g$^{-1}$ |
| Mean radius $X_0$ (SASfit) | $(8.52 \pm 0.04)$ nm | $(37.65 \pm 3.30)$ nm |
| Mean radius (certification report)[b] | $(9.1 \pm 0.8)$ nm | $(42.0 \pm 1.1)$ nm |
| Width of distribution $s$ (nm) | $(2.00 \pm 0.03)$ nm | $(8.29 \pm 3.04)$ nm |
| Number ratio $N_1/N_2$ | 1567 ± 371 (± 24 %) | |
| Mass ratio $\phi_{m,1}/\phi_{m,2}$ (SASfit) | 18.5 ± 4.6 (± 25 %) | |
| $\phi_{m,1}/\phi_{m,2}$ (certification report) | 19.8 | |

[a]Values were calculated from the information on the production data given in the certification report of ERM-FD-102 (Kestens & Roebben, 2014). [b]Number-weighted modal area-equivalent diameter as obtained by transmission and scanning electron microscopy (Kestens & Roebben, 2014).

## 7. Documentation

A comprehensive manual is included in the delivery package. It contains the physical and mathematical details and definitions of the internal algorithms as well as documentation and implementation notes for most of the models. Additionally, there is a collaborative wiki website available containing further information on installation details and providing help for setting up and writing custom plug-ins. For core topics in using the SASfit program there are also video guides available online (Youtube.com, 2014). The numerical part of SASfit is written in C and the user interface in Tcl/Tk. The latest packages of version 0.94.6 are available at http://sf.net/projects/sasfit/files/0.94.6.

## 8. Outline

On top of the elaborated developments for general use presented here, recently there has also been made an extensive development effort in order to implement a new solver for the Ornstein-Zernike equation for different closure relations and potentials. A specialized user interface plots the numerical solution which can be used for structure factor input in model dependent analysis. The details of the solver and its implementation as well as its usage will be presented in another publication elsewhere. Starting with the early versions of the SASfit program, it uses the Levenberg-Marquardt (Levenberg, 1944) algorithm to find solutions for multi-dimensional non-linear optimization problems, which are posed by small-angle scattering data analysis. Users often experience stability issues or sometimes even crashes of the optimization routine of SASfit especially when it comes to optimizing several parameters of a complex model at once. Those issues may be caused by correlated parameters within a model, but it is rarely possible to predict those circumstances. A future goal in further development of the SASfit software is to provide more optimization algorithms to the user. There are modern versions of the Levenberg-Marquardt-Algorithm available, which are expected to exhibit improved numerical stability over the old implementation. Also allowing parameter constrains would help to stabilize the fit. At the moment the fit is aborted, if a parameter is running out of its defined range. Better minimization routines may be able to automatically account for that. This might improve the overall workflow and user experience with the SASfit analysis program.

## References

De Temmerman, P. J., Verleysen, E., Lammertyn, J. & Mast, J. (2014). *Journal of Nanoparticle Research* **16**, 2628.
Finsy, R., Moreels, E., Bottger, A. & Lekkerkerker, H. (1985). *Journal of Chemical Physics* **82**, 3812-3816.
Forster, S., Apostol, L. & Bras, W. (2010). *Journal of Applied Crystallography* **43**, 639-646.
Galassi, M. & Gough, B. (2009). *Gnu scientific library: Reference manual*, 3 ed. Network Theory.
Gleber, G., Cibik, L., Haas, S., Hoell, A., Müller, P. & Krumrey, M. (2010). *Journal of Physics: Conference Series* **247**, 012027.
Hamilton, W. C. (1965). *Acta Crystallographica* **18**, 502-510.
Hammouda, B. (2012). *Macromolecular Theory and Simulations* **21**, 372-381.
http://sourceforge.net/projects/sasfit/
Ilavsky, J. & Jemian, P. R. (2009). *Journal of Applied Crystallography* **42**, 347-353.
IUCr (2008). *R factor,* http://reference.iucr.org/dictionary/R_factor.
Kaiser, D. L. & Watters, R. L. (2007). National Institute of Standards & Technology.
Kestens, V. & Roebben, G. (2014). European Commission, Joint Research Centre, Institute for Reference Materials and Measurements (IRMM).
Levenberg, K. (1944). *Quaterly of Applied Mathematics* **2**, 164-168.
Meli, F., Klein, T., Buhr, E., Frase, C. G., Gleber, G., Krumrey, M., Duta, A., Duta, S., Korpelainen, V., Bellotti, R., Picotto, G. B., Boyd, R. D. & Cuenat, A. (2012). *Measurement Science & Technology* **23**, 125005.
Metrology, J. J. C. f. G. i. (2008). Evaluation of measurement data - guide to the expression of uncertainty in measurement (gum), Vol. 100: Joint Committe for Guides in Metrology (JCGM).
Mylonas, E. & Svergun, D. I. (2007). *Journal of Applied Crystallography* **40**, S245-S249.
Orthaber, D., Bergmann, A. & Glatter, O. (2000). *Journal of Applied Crystallography* **33**, 218-225.
Pauw, B. R., Pedersen, J. S., Tardif, S., Takata, M. & Iversen, B. B. (2013). *Journal of Applied Crystallography* **46**, 365-371.
Petoukhov, M. V., Franke, D., Shkumatov, A. V., Tria, G., Kikhney, A. G., Gajda, M., Gorba, C., Mertens, H. D. T., Konarev, P. V. & Svergun, D. I. (2012). *Journal of Applied Crystallography* **45**, 342-350.
Potočnik, J. (2011). Commission recommendation of 18 october 2011 on the definition of nanomaterial text with eea relevance edited by E. Commision, pp. 38-40. Brussels: European Commision.
Sourceforge.net (2014). *Download statistics,* http://sourceforge.net/projects/sasfit/files/stats/timeline?dates=2014-01-01+to+2014-12-31.
Varga, Z., Yuana, Y., Grootemaat, A. E., van der Pol, E., Gollwitzer, C., Krumrey, M. & Nieuwland, R. (2014). *Journal of extracellular vesicles* **3**, 23298.
Youtube.com (2014). *Sasfit user guides,* https://www.youtube.com/user/SASfitTeam.